\documentclass[pre,twocolumn]{revtex4b5}
\usepackage{epsfig, amssymb, hyperref}

\begin{document}
\title{Variational Method for Calculation of Plasma Phase Diagrams in Path Integral Representation}
\author{Ilmars Madzhulis}
\author{Vilnis Frishfelds}
\email{mf60006@lanet.lv}
\affiliation{Faculty of Physics and Mathematics, University of Latvia, Zellu 8, Riga LV-1002,
Latvia}

\begin{abstract}
The use of variational method in functional integral approach is discussed for fermion and boson systems with Coulomb
interaction. The formal general expression of thermodynamic potential is obtained by Feynman path integral technique and
representation of Coulomb interaction with functional integrals. Introduced additional complex field show to transform the
problem to calculation of functional integrals containing third order vertices. The thermodynamic potential can be found
from variational principle with respect to field cumulants. The calculation of the equation of state and critical properties
is demonstrated for symmetrical plasma by variation of finite number of parameters in the propagator.
\end{abstract}

\maketitle

\section{Introduction}
The approach of functional integrals is used more and more frequently studying the properties of non-relativistic Coulomb
plasmas. The Feynmann path integral technique allows obtaining the properties of plasmas in systematic manner without
utilising such approximation as chemical picture or Pad\'{e} formulaes. The path integral method takes the advantages of
Green function formalism. Such \textit{ab initio} method as restricted path integral Monte Carlo simulations \cite{Ceperley}
provide plausible results at high temperature, where the experiments allows only some implicit measurements. The further
development of these simulations is prospective for higher number of particles. The direct path integral method by Filinov
\cite{Filinov} rigorously includes anti-symmetrisation but uses the effective pair potential. Both methods are well suited
for high temperature calculations. However, these simulations are still time consuming for complex fermionic systems.
Therefore, the deeper understanding of path integral approach should be reached. The Hubbard-Schofield transformation has
been a great leap forward for functional methods \cite{Siegert}. Recently, Brown and Yaffe \cite{Brown} have reported that
additional integral over complex field is commendable calculating the action of a charged particle. However, the approach
in principe cannot be used for highly degenerate plasmas. The paper will show how the anti-symmetrisation effects could be
included accurately by introduction of additional integration over complex field for large canonical ensemble. Tough, the
mathematical methods to calculate the obtained formal general expression of partition function containing functional
integrals are insufficiently effective. There exists a variational approach using the cumulative averages of fields for
system in thermodynamic equilibrium. A universal function of field cumulants that does not depend on physical model should
be known for this purpose. The comparatively inaccurate expression of the required function is obtained using the simplest
diagram in expansion. The variational method is well suited for qualitative study of complex plasmas in the same manner as
"Gaussian" packet in quantum mechanics choosing the appropriate class of variable functions.
\section{Anti-symmetrisation}
The Coulomb interaction belongs to a class with pair and positively defined potentials
\begin{equation}
 V(x)=\frac{1}{V}\sum_\mathbf{k}V_k e^{i\mathbf{kx}},\quad V_k >0.
\end{equation}
For such kind of potentials, the Hubbard-Schofield transformation allows to write the density matrix of the system of $n$
non-identical particles as \cite{Skrypnik}
\begin{widetext}
\begin{equation}\label{int6}
 \rho_n(\mathbf{X},\mathbf{X}')=\int\mathcal{D}\varphi(\mathbf{x},u)\exp\left[-\frac{1}{\hbar}\int_0^U
 du\sum_\mathbf{k}\frac{|\varphi_\mathbf{k}(u)|^2}{2V_k}\right]
 \exp\left[\frac{UV(0)}{2\hbar}\sum_{j=1}^n q_j^2\right] \prod_{j=1}^n\rho_j(\mathbf{x}_j,\mathbf{x}'_j,U),
\end{equation}
\begin{equation}\label{gamma}
 \langle
 \mathbf{x}|\hat{\rho}_j(U)|\mathbf{x'}\rangle=\rho_j(\mathbf{x},\mathbf{x'},U)
 =\int_\mathbf{x}^\mathbf{x'}\mathcal{D}\mathbf{x}(u)\exp\left[-\frac{1}{\hbar}\int_0^Udu\left(\frac{m_j\dot{\mathbf{x}}^2(u)}{2}+i
q_j \varphi(\mathbf{x}(u))\right)\right].
\end{equation}
\end{widetext}
where integration is carried out over all configurations of the real field $\varphi(\mathbf{x},u)$ in 4-dimensional space
$\mathbf{x}\in V$, $u\in [0,U\equiv\hbar\beta]$;
$\mathbf{X}(u)=\{\mathbf{x}_1(u),\mathbf{x}_2(u),\ldots,\mathbf{x}_n(u)\}$; $q_j$ - a charge of particle;
$\varphi_\mathbf{k}$ is the Fourier component of electric field
\begin{equation}
  \varphi(\mathbf{x})=\frac{1}{\sqrt{V}}\sum_\mathbf{k}\varphi_\mathbf{k}e^{i\mathbf{kx}}.
\end{equation}
The integration over real field $\varphi(\mathbf{x},u)$ does not depend on the number of particles, their charge and type
of symmetry, Therefore, matrix $\rho_j(\mathbf{x},\mathbf{x}',U)$ can be considered formally as density matrix of
noninteracting particles placed in an external field $i\varphi(\mathbf{x},u)$. The Coulomb interaction has been taken into
account by integration over all configurations of this field. The result is similar to the Feynman interpretation of
quantum electrodynamics by path integrals \cite{FeynPaIn}, except the exclusion of the self-interaction part $V(0)$ in
(\ref{int6}). If some of the particles are identical, then either symmetry or anti-symmetry must be taken into account by
summation over all permutations. The symmetrisation procedure should be performed for the system of noninteracting
particles, according to the statement above. We are interested in the equilibrium properties of the system of charged
particles, such as equation of state or phase diagram. After symmetrisation \cite{FeynStPh}, the partition function of
large canonical ensemble in path integral representation becomes
\begin{eqnarray}\label{primstatsum}
 \mathcal{Z}=\int\mathcal{D}\varphi(\mathbf{x},u)\exp\left[-\frac{1}{\hbar}\sum_\mathbf{k}\int_0^Udu\frac{|\varphi_\mathbf{k}(u)|^2}{2V_k}\right.\nonumber\\
 \left.-\sum_j\zeta_j \mathrm{Tr}\ln(\hat{1}-\zeta_j e^{\beta\mu_j^{tot}} \hat{\rho}_j(U))\right],
\end{eqnarray}
where $\langle\mathbf{x}|\hat{1}|\mathbf{x}'\rangle=\delta(\mathbf{x}-\mathbf{x}')$; $j$ now is an index of particle
species (including different spin orientations); $\zeta_j=\pm1$ for bosons or fermions, respectively; $\mu_j^{tot}$ -
the chemical potential of $j$-type charged particle. The Fourier component $V_k$ of modified Coulomb potential $V(x)$
is
\begin{eqnarray}\label{coulomb}
 V(x)=\frac{1}{x}(1-e^{-\Upsilon x}),\quad V_k=\frac{4\pi}{k^2+\frac{k^4}{\Upsilon^2}},
\end{eqnarray}
where a large parameter $\Upsilon$ is introduced to avoid from the singularity of Coulomb potential at small distances. The
constant self-interaction part (\ref{int6}) can be added to chemical potential
\begin{equation}\label{remu}
 \mu_j^{tot}=\mu_j+\frac{q_j^2V(0)}{2},
\end{equation}
where $\mu_j$ is the conventional chemical potential of the system. At the limit $\Upsilon\rightarrow\infty$, the terms
proportional to $\Upsilon$ in the partition function must contract. The short distance divergences can be excluded also by
introduction of fractional dimension \cite{Brown}.

\section{Variational principle for cumulants}\label{ParFun}
In the following section, the basic tool will be constructed that helps to find the thermodynamic potential from
variational principle for cumulative averages (cumulants) of some field. It is simply the method how to calculate the
non-Gaussian type integrals. The idea was applied for the study of the phase transitions in Landau-Ginzburg theory
\cite{Madzhulis}. Consider the partition function as a one-dimensional integral
\begin{equation}\label{v1}
 \mathcal{Z}=\int_{-\infty}^\infty d\varphi e^{Q(\varphi)} .
\end{equation}
The statistical average of $\delta$-function is
\begin{equation}\label{v2}
 \langle\delta(f-\varphi)\rangle_f=\frac{1}{\mathcal{Z}} \int_{-\infty}^\infty
 df\delta(f-\varphi)e^{Q(f)}=\frac{e^{Q(\varphi)}}{\mathcal{Z}}.
\end{equation}
Therefore, the term $Q(\varphi)$ which depends on the physical model can be isolated in partition function
\begin{equation}\label{v3}
  F\equiv \ln\mathcal{Z}=Q(\varphi)-\ln\langle\delta(f-\varphi)\rangle_f.
\end{equation}
The partition function cannot depend on variable $\varphi$. Hence, it equals to its average:
\begin{equation}\label{v4}
  F=\langle Q(\varphi)\rangle_\varphi-\langle\ln\langle\delta(f-\varphi)\rangle_f\rangle_\varphi.
\end{equation}
We can expand the average of $\delta$-function in terms of cumulants using the integral representation of
$\delta$-function
\begin{eqnarray}\label{v5}
  \langle\delta(f)\rangle_f=\frac{1}{2\pi}\int_{-\infty}^\infty dh\langle\exp[i h
  f]\rangle_f\nonumber\\
  =\frac{1}{2\pi}\int_{-\infty}^\infty dh\exp\left[\sum_{n=1}^\infty\frac{(i
  h)^n}{n!}K_n^f\right],
\end{eqnarray}
where $K_n^f=\langle f^n\rangle_c$ is the $n$-th order cumulant of the variable $f$. The relation to the usual average
$\langle f^n\rangle$,
\begin{equation}\label{v6}
 \sum_{n=0}^\infty\frac{(i h)^n}{n!}\langle f^n\rangle =\exp\left[\sum_{n=1}^\infty\frac{(i h)^n}{n!}K_n^f\right]\equiv P,
\end{equation}
is utilised in (\ref{v5}). Let us denote the $n$-th order cumulant of variable $\varphi$ as $K_n^\varphi=\langle
\varphi^n\rangle_c$. The variation of function $F$ with respect to the total cumulant $K_n$ consists of two parts
\begin{equation}\label{v7}
  \frac{\partial F}{\partial K_n}=\frac{\partial F}{\partial K_n^f} +\frac{\partial F}{\partial K_n^\varphi}.
\end{equation}
It follows from the $F$ independence of $\varphi$ that $\frac{\partial F}{\partial K_n^\varphi}=0$. The derivative in
respect to the cumulant $K_n^f$, in compliance with (\ref{v2}, \ref{v5}, \ref{v6}) is zero, too:
\begin{eqnarray}\label{v8}
 \frac{\partial F}{\partial K_n^f}&=&\left\langle\frac{\partial F}{\partial K_n^f}\right\rangle_\varphi\nonumber\\
 &=&\left\langle \frac{1}{2\pi\langle\delta(f-\varphi)\rangle_f}\int_{-\infty}^\infty dh\frac{\partial P}{\partial K_n^f} e^{-i h\varphi}\right\rangle_\varphi\nonumber\\
&=&\frac{\mathcal{Z}}{2\pi}\int_{-\infty}^\infty dh\frac{\partial P}{\partial K_n^f} \left\langle e^{-Q(\varphi)-i h\varphi}
\right\rangle_\varphi\\
&=&\int_{-\infty}^\infty dh\frac{\partial P}{\partial K_n^f} \delta(h)=0\nonumber.
\end{eqnarray}
Subsequently, the variational principle is valid for thermodynamic potential with respect to cumulants $K_n$. The average
$\langle Q(\varphi)\rangle_\varphi$ depends only on physical model and usually includes a finite number of $K_n$. On the
other hand, the average of the second term in (\ref{v4}) expressed by cumulants does not depend on interaction parameters,
that makes the analogy of this term with an entropy. However, such universal expression is hardly obtainable. The expansion
in terms of cumulants will be shown in next section using a diagram technique.
 \\If the partition function is the functional integral of some field $\varphi(\mathbf{x})$ in a box $\mathbf{x}\in V$
then it contains a product of $\delta$-functions
\begin{equation}\label{v10}
 F=\langle Q[\varphi] \rangle_\varphi-\left\langle\ln\left\langle\prod_\mathbf{k}\delta(f_\mathbf{k}-\varphi_\mathbf{k})\right\rangle_f\right\rangle_\varphi
\end{equation}
in Fourier representation. Henceforth, the partition function involves mixed cumulative averages $\langle
\varphi_\mathbf{k_1}\varphi_\mathbf{k_2} \ldots \rangle_\varphi$ of physically different modes. But for a homogeneous
system the part of interaction $Q[\varphi]$ does not include mixed cumulants of $\varphi_\mathbf{k}$. Thus, one root after
variation is always zero.

\section{Diagram expansion on field cumulants}
The diagram technique is discussed in a number of papers both for classical plasma and quantum plasmas, e.g.,
\cite{Ortner}, \cite{Alastuey}. The expansion is usually made in terms of Debye screening radius. Such an approach is
commendable for strict expansion at low densities. However, we need a specific diagram expansion on the cumulants in order
to apply the variational principle. In first approximation, we will use the expansion in diagrams up to the square of
charge. The final thermodynamic potential will likewise contain the charge of particle up to its square. Of course, we will
obtain similar results as in mentioned papers though with different mathematical approach. Variational principle provides
essential advantage in comparison with virial or activity expansion studying the phase transitions in plasma. To apply the
variational principle it is necessary to express the last term of (\ref{v4}) in cumulants of some field $\varphi$. A general
expansion exists probably for this quite universal mathematical task, but here the diagram technique is used for this
purpose. The integral representation of the average of $\delta$-function is (see (\ref{v5}))
\begin{eqnarray}\label{dt2}
 &&\left\langle\delta(f-\varphi)\right\rangle_f\nonumber\\
 &&=\frac{1}{2\pi}\int_{-\infty}^\infty dh\exp\left[\sum_{n=1}^\infty\frac{(i h)^n}{n!}K_n-i h\varphi\right],
\end{eqnarray}
where $K_n=\langle f^n\rangle_c=\langle \varphi^n\rangle_c$ is the $n$-th order cumulant. Let us denote the cumulant $K_2$
of the Gaussian part as $G$. The logarithm of (\ref{dt2}) can be formally expressed by means of diagrams as follows
\begin{widetext}
\begin{eqnarray}
 \ln\left\langle\delta(f-\varphi)\right\rangle_f=\ln\left\{\frac{1}{2\pi}\int_{-\infty}^\infty dh\exp\left[-\frac{h^2G}{2}+\sum_{n\neq 2}\frac{(i h)^n}{n!}K_n-i h\varphi\right]\right\}\nonumber\\
 =-\frac{1}{2}\ln(2\pi G)-\frac{(K_1-\varphi)^2}{2G}+\quad\raisebox{-7pt}{\epsfig{file=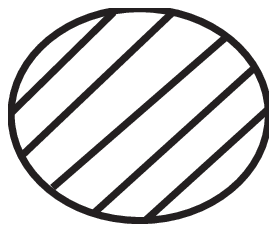, height=0.7cm}}\quad+\sum_{n=1}^\infty\frac{(-i\varphi)^n}{n!}\quad \raisebox{-11pt}{\epsfig{file=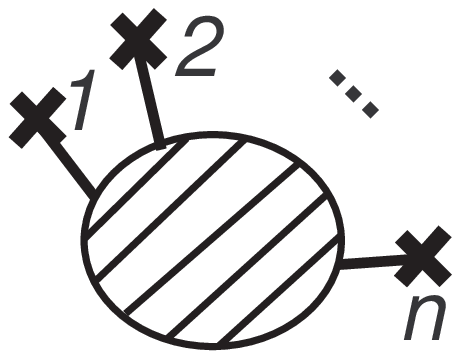, height=1.2cm}}\quad\label{dt5},
\end{eqnarray}
\end{widetext}
where the first two terms follow from the Gaussian approximation; the first graph represents the sum of all joint diagrams
that does not contain $\varphi$, but the second graph all joint diagrams containing $\varphi$. We need to find an average
of (\ref{dt5}) with respect to $\varphi$. The average $\langle\varphi^n\rangle$ can be written by cumulants in accordance
with Vick's theorem
\begin{equation}\label{dt7}
 \left\langle\varphi^n\right\rangle=\left\langle\varphi^n\right\rangle_c+n\langle\varphi\rangle_c\left\langle
 \varphi^{n-1}\right\rangle_c+\dots\quad.
\end{equation}
Therefore, the averaging of (\ref{dt5}) couples the first order vertices $\mathbf{\times}$ to produce higher order
vertices. The resulting diagrams may cancel with the similar diagrams in the first graph of (\ref{dt5}). Only certain kind
of diagrams survives. Thus, all diagrams that contain first order vertices or vertices joining vertex to itself (loop)
cancels. As a result, the average of (\ref{dt5}) does not contain the cumulant $K_1=\langle\varphi\rangle$. The simplest
remaining diagram consists of two $K_3$ vertices:
\begin{eqnarray}\label{dt10}
 -\frac{1}{2}
 \raisebox{3pt}
  {
  \begin{picture}(30,0)
  \thicklines
  \put(15,0){\circle{20}}
  \put(5,0){\circle*{5}}
  \put(25,0){\circle*{5}}
  \put(5,0){\line(25,0){20}}
 \end{picture}
 }
 =\frac{K_3^2}{2G^3}.
\end{eqnarray}
All three edges cannot be crossed simultaneously, since the diagram would split apart before averaging with respect to
$\varphi$. Therefore, the first few terms in expansion are
\begin{equation}\label{dt11}
 \left\langle\ln\left\langle\delta(f-\varphi)\right\rangle_f\right\rangle_\varphi=-\frac{1}{2}\ln(2\pi G)-\frac{1}{2}+\frac{K_3^2}{2G^3}.
\end{equation}
The intrinsic energy part in Fourier representation usually does not contain cumulative averages of order higher than
second. Then, the extremum  condition $\frac{\partial\ln\mathcal{Z}}{\partial K_3}=0$ yields $K_3=0$. However, that could
probably violate for additional extremums if the next diagrams are considered.

\section{Complex field for fermions}\label{cmplxfld}
The symmetry properties of particles are accounted in the partition function via the logarithmic term in
(\ref{primstatsum}). It seems that such a construction spoils its solvability due to the slow convergence of Taylor series
for strongly coupled plasma. Nevertheless, let us remember that any nonsingular quadratic matrix satisfies equation
$\mathrm{Tr}\ln K=\ln\det K$. At the same time, the determinant of matrix $K$ can be represented as the path integral over
all configurations of complex field $\Psi(\mathbf{x})$ \cite{Vasiljev}:
\begin{eqnarray}\label{det}
 &&\frac{1}{\det K}\\
 &\sim&\int\mathcal{D}\Psi(\mathbf{x})\exp\left[-\int d\mathbf{x}\int d\mathbf{x}'\Psi(\mathbf{x}) K(\mathbf{x},\mathbf{x}')\Psi^*(\mathbf{x}')\right]\nonumber.
\end{eqnarray}
This leads to the idea that the logarithmic part of the partition function (\ref{primstatsum}) for fermions can be
represented as a functional integral
\begin{eqnarray}\label{t1}
  &&\exp[-\mathrm{Tr}\ln(\hat{1}+e^{\beta\mu_j^{tot}} \hat{\rho}_j(U))]\nonumber\\
  &=&\int\mathcal{D}\Psi_j(\mathbf{x})\exp\left[-\int d\mathbf{x} |\Psi_j(\mathbf{x})|^2\right.\\
  &&\left.+e^{\beta\mu_j^{tot}+i\pi}\int d\mathbf{x}\int d\mathbf{x'}
 \Psi_j(\mathbf{x})\rho_j(\mathbf{x},\mathbf{x'},U)\Psi_j^*(\mathbf{x'})\right]\nonumber.
\end{eqnarray}
The multiplier $e^{i\pi}$ is added in (\ref{t1}) to change the sign for fermions. The change of the sign is unnecessary for
bosons: the difference will be shown latter. Note an analogy of function $\Psi_j(\mathbf{x})$ with the stationary wave
function in quantum mechanics. As mentioned above, the integration over complex variables $\Psi$ is introduced also in
\cite{Brown} for canonical ensemble. After splitting the imaginary-time interval in $N\gg 1$ small parts
$\varepsilon=\frac{U}{N}$, the kernel $\rho_j(\mathbf{x},\mathbf{x'},U)$ (\ref{gamma}) of linear integral operator
$\hat{\rho}_j$ becomes a convolution \cite{FeynPaIn}
\begin{widetext}
 $$
 \rho_j(\mathbf{x},\mathbf{x'},U)
=\int d\mathbf{x}_{N-1}\ldots\int d\mathbf{x}_2\int d\mathbf{x}_1\rho_j(\mathbf{x},\mathbf{x}_{N-1},\varepsilon)\ldots
\rho_j(\mathbf{x}_{2},\mathbf{x}_{1},\varepsilon) \rho_j(\mathbf{x}_1,\mathbf{x}',\varepsilon).
 $$
Henceforth, the logarithmic part (\ref{t1}) of the partition function can be transformed to a Gaussian type functional
integral over all configurations of complex field $\Psi_j(\mathbf{x},u)$ in 4-dimensional space \cite{Vasiljev}
\begin{eqnarray}\label{t2d}
  \exp[-\mathrm{Tr}\ln(\hat{1}+e^{\beta\mu_j^{tot}} \hat{\rho}_j(U))]
=\int\mathcal{D}\Psi_j(\mathbf{x},u)\exp\left[-\sum_u\int d\mathbf{x}|\Psi_j(\mathbf{x},u)|^2\right.\nonumber\\
 + \left. \sum_u\int d\mathbf{x}\int d\mathbf{x'}\Psi_j(\mathbf{x},u+\varepsilon)\exp\left[\frac{\varepsilon(\beta\mu_j^{tot}+i\pi)}{U}\right]\rho_j(\mathbf{x},\mathbf{x}',\varepsilon)    \Psi_j^*(\mathbf{x'},u)\right],\nonumber\\
 u=0,\varepsilon,\ldots,(N-1)\varepsilon.\nonumber
\end{eqnarray}
\end{widetext}
For Coulomb system with density matrix $\rho_j(\mathbf{x},\mathbf{x'},\varepsilon)$ (\ref{gamma}), the charged particle can
be considered as placed in imaginary field $i\varphi$, their Hamiltonian being
$\hat{H}_j=-\frac{\hbar^2}{2m_j}\Delta_\mathbf{x}+i q_j\varphi$ (a non-Hermitian operator). The statistical density matrix
$\rho_j(\mathbf{x},\mathbf{x'},u)$ satisfies the equation $\hbar\frac{\partial\rho_j}{\partial u}=-\hat{H}_j\rho_j$
analogous to Schr\"{o}dinger equation. The expansion of this density matrix and small $\varepsilon$ is
$\rho_j(\mathbf{x},\mathbf{x'},\varepsilon)$ for small $\varepsilon$ is
\begin{eqnarray}\label{t2}
 \rho_j(\mathbf{x},\mathbf{x'},\varepsilon)\approx\left(1+\frac{\varepsilon\hbar}{2 m_j}\Delta_\mathbf{x}-\frac{\varepsilon}{\hbar}i q_j\varphi(\mathbf{x'},u)\right.\nonumber\\
 \left.-\frac{\varepsilon^2}{2\hbar^2} q_j^2\varphi^2(\mathbf{x'},u) \right)\delta(\mathbf{x}-\mathbf{x'}).
\end{eqnarray}
Hence, the partition function (\ref{primstatsum}) becomes a functional integral over all configurations of the real field
$\varphi(\mathbf{x},u)$ and the complex field $\Psi_j(\mathbf{x},u)$
\begin{widetext}
\begin{equation}\label{t5}
  \mathcal{Z}=\int\mathcal{D}\varphi(\mathbf{x},u)
 \frac{\exp\left[-\frac{\varepsilon}{\hbar}\sum_{\mathbf{k},u}\frac{|\varphi_\mathbf{k}(u)|^2}{2V_k}\right]}{\prod_j\int\mathcal{D}\Psi_j(\mathbf{x},u)\exp\left[\sum_u\int d\mathbf{x}[\Psi_j(\mathbf{x},u+\varepsilon)-\Psi_j(\mathbf{x},u)-\frac{\varepsilon}{\hbar}\Psi_j(\mathbf{x},u+\varepsilon)\hat{\mathcal{H}}_j ]\Psi_j^*(\mathbf{x},u) \right]},
\end{equation}
\begin{equation}
 \quad\quad\hat{\mathcal{H}}_j=-\frac{\hbar^2}{2m_j}\Delta_\mathbf{x}+i
  q_j\varphi(\mathbf{x},u)+\frac{\varepsilon}{2\hbar}q_j^2\varphi^2(\mathbf{x},u) -\mu_j^{tot}-\frac{i\pi\hbar}{U}.\label{t6}
\end{equation}
\end{widetext}
The number of complex fields coincides with the number of particle species. The Fourier transformation of these fields is
useful:
\begin{eqnarray}\label{t7}
  \varphi(\mathbf{x},u)=\frac{1}{\sqrt{V\beta}}\sum_{\mathbf{k},\omega}\varphi_{\mathbf{k},\omega}e^{i
  \mathbf{k}\mathbf{x}+i\omega u},\nonumber\\
  \Psi_j(\mathbf{x},u)=\frac{1}{\sqrt{V}}\sum_{\mathbf{k},\omega}\Psi_{j,\mathbf{k},\omega}e^{i
  \mathbf{k}\mathbf{x}+i\omega u},\nonumber\\
  \omega=\frac{2\pi n}{U},\quad
  n=-M,-M+1,\ldots,M\label{t7b}
\end{eqnarray}
assuming for convenience that $N=2M+1$ is an odd integer. The $\varepsilon$ term in operator $\hat{\mathcal{H}}_j$
(\ref{t6}) should be neglected, but for systems with Coulomb interaction it survives and accounts for self-interaction. As
we shall see in (\ref{extrem}), the average $G_{\mathbf{k},\omega}\equiv\langle|\varphi_{\mathbf{k},\omega}|^2\rangle$ at
$|\omega|\rightarrow\infty$ equal to the interaction potential $V_k$. Thereafter, the self-interaction part arises,
according to the Fourier expansion of Coulomb interaction (\ref{coulomb}):
\begin{equation}\label{t7a}
 \lim_{\varepsilon\rightarrow 0}\varepsilon\langle\varphi^2(\mathbf{x},u)\rangle=\lim_{\varepsilon\rightarrow 0}\frac{\varepsilon N}{V\beta}\sum_\mathbf{k}V_k=\hbar V(0).
\end{equation}
The self-interaction part cancels if the chemical potential of Coulomb system $\mu_j^{tot}$, (\ref{remu}), is subtracted in
(\ref{t6}). The partition function (\ref{t5}) for small $\varepsilon$ becomes
\begin{widetext}
\begin{equation}\label{t10}
 \mathcal{Z}=\int\mathcal{D}\varphi(\mathbf{x},u)
\frac{\exp\left[-\sum_{\mathbf{k},\omega}\frac{|\varphi_{\mathbf{k},\omega}|^2}{2V_k}
\right]}{\prod_j\int\mathcal{D}\Psi_j(\mathbf{x},u)\exp\left[-\sum_{\mathbf{k},\omega}\left(\alpha_{j,\mathbf{k},\omega}|\Psi_{j,\mathbf{k},\omega}|^2+\frac{i
q_j\sqrt{\beta}\varphi_{\mathbf{k},\omega}}{\sqrt{V}}
\sum_{\mathbf{q},\chi}\Psi_{j,\mathbf{q},\chi}\Psi_{j,\mathbf{k}+\mathbf{q},\omega+\chi}^*\right)\right]},\nonumber
\end{equation}
\end{widetext}
where
\begin{eqnarray}\label{t11}
 \alpha_{j,\mathbf{k},\omega} =N(e^{i\omega\varepsilon}-1 )-i\pi+c_{j,\mathbf{k}}^{id},\\
 c_{j,\mathbf{k}}^{id}=\beta\left(\frac{\hbar^2\mathbf{k}^2}{2m_j}-\mu_j\right)\nonumber.
\end{eqnarray}
The term $-i\pi$ is not present for the system of bosons. Moreover, the integration over complex field
$\Psi_j(\mathbf{x},u)$ in partition function $\mathcal{Z}$ for bosons is present in numerator.

\section{Variational principle for fermions}
According to the previous section, the following integral is to be found in quantum statistics for Fermi systems:
\begin{equation}\label{f1}
  \mathcal{Z}=\int d\varphi\frac{\exp[Q(\varphi)]}{\int d\Psi\exp[\Phi(\varphi,|\Psi|^2)]},
\end{equation}
where $d\Psi=d\mathrm{Re}\Psi d\mathrm{Im}\Psi$; $\varphi$ is real variable, but $\Psi$ - complex. The average values of
$\delta$-function are
\begin{eqnarray}
  &&\langle\delta(f-\varphi)\delta(\Psi-\Theta)\rangle_{f,\Psi}\nonumber\\
  &=&\frac{1}{\mathcal{Z}}\frac{\exp[Q(\varphi)]\exp[\Phi(\varphi,|\Theta|^2)]}{\left(\int
  d\Psi_1\exp[\Phi(\varphi,|\Psi_1|^2)]\right)^2}\label{f4},
\end{eqnarray}
\begin{equation}
   \langle\delta(f-\varphi)\rangle_f=\frac{1}{\mathcal{Z}}\frac{\exp[Q(\varphi)]}{\int
  d\Psi_1\exp[\Phi(\varphi,|\Psi_1|^2)]}\label{f6}.
\end{equation}
Therefore, the partition function can be split using $\delta$-functions (\ref{f4}, \ref{f6}) in analogy with simple field
of variable $\varphi$ in (\ref{v4})
\begin{eqnarray}
  \ln\mathcal{Z}&=&\langle Q(\varphi)\rangle_\varphi-\langle\Phi(\varphi,|\Theta|^2)\rangle_{\varphi,\Theta}\nonumber\\
  &-&2\langle\ln\langle\delta(f-\varphi)\rangle_f\rangle_\varphi\nonumber\\
  &+&\left\langle\ln\langle\delta(f-\varphi)\delta(\Psi-\Theta)\rangle_{f,\Psi}\right\rangle_{\varphi,\Theta}\nonumber .
\end{eqnarray}
The conclusions are similar to those made in sect.~\ref{ParFun}, i.e., the variation of partition function with respect to
cumulants $\langle f^l\Psi^m\Psi^{*n}\rangle_c$ must be zero. In our case, we should consider the fields in
coordinate-temperature space, $\varphi(\mathbf{x},u)$, $\Psi_j(\mathbf{x},u)$ as in (\ref{v10}). The first two terms follow
from functional integral representation (\ref{t10}):
\begin{eqnarray}
 \langle Q[\varphi]\rangle_\varphi-\langle\Phi[\varphi,|\Theta|^2]\rangle_{\varphi,\Theta}=-\sum_{\mathbf{k},\omega}\frac{\langle|\varphi_{\mathbf{k},\omega}|^2\rangle_\varphi}{2V_k}\nonumber\\
 +\sum_{j,\mathbf{k},\omega} \alpha_{j,\mathbf{k},\omega}\langle|\Theta_{j,\mathbf{k},\omega}|^2\rangle_{\Theta}\label{f12}\\
 +\frac{i \sqrt{\beta}}{\sqrt{V}}\sum_{j,\mathbf{k},\omega,\mathbf{q},\chi} q_j\left\langle\varphi_{\mathbf{k},\omega}
\Theta_{j,\mathbf{q},\chi}\Theta_{j,\mathbf{k}+\mathbf{q},\omega+\chi}^*\right\rangle_{\varphi,\Theta}\nonumber.
\end{eqnarray}
One recognises here the terms representing the energy of electric field, the energy of noninteracting fermions, and the
particle-field interaction characterised by the third order vertex $\langle \varphi(\mathbf{x},u) \Psi_j(\mathbf{x},u)
\Psi_j^*(\mathbf{x},u) \rangle$ in the coordinate-temperature representation.

\section{Diagram expansion on cumulants for fermions}
The expansion of
\begin{eqnarray}
 \langle\ln\langle\delta(f-\varphi)\delta(\Psi-\Theta)\delta(\Psi^*-\Theta^*)\rangle_{f,\Psi,\Psi^*}\rangle_{\varphi,\Theta,\Theta^*},\\
 f,\varphi\in \mathbb{R},\quad \Psi,\Theta\in \mathbb{C}\nonumber
\end{eqnarray}
in terms of cumulants is necessary for fermion plasma. By using the same scheme as in the previous section, the integral
representation of $\delta$-function gives an expansion in cumulative averages
\begin{widetext}
\begin{eqnarray}
 &&\langle\delta(f-\varphi)\delta(\Psi-\Theta)\delta(\Psi^*-\Theta^*)\rangle_{f,\Psi,\Psi^*}\nonumber\\
&=&\frac{1}{(2\pi)^3}\int_{-\infty}^\infty \int_{-\infty}^\infty \int_{-\infty}^\infty dhdadb\exp\left[\sum_{n_1+n_2+n_3\geq
1}\frac{(i h)^{n_1}(i A)^{n_2}(i A^*)^{n_3}}{n_1!n_2!n_3!} \left\langle f^{n_1}\Psi^{n_2}\Psi^{*n_3}\right\rangle_c-i
h\varphi-i A\Theta-i A^*\Theta^*\right],\nonumber
\end{eqnarray}
\end{widetext}
where $a=\mathrm{Re} A$ and $b=\mathrm{Im} A$ are real integration variables. Since $\Psi$ and $\Psi^*$ are complex-valued,
the cumulants $\langle f^{n_1}\Psi^{n_2}\Psi^{*n_3}\rangle_c$ are zero if $n_2$ differs from $n_3$. Moreover, the average
$\langle\varphi\rangle$ is zero for neutral plasma. Therefore, the first nonzero summands of the sum in previous equation
are
\begin{eqnarray}
 \sum_{n_1+n_2+n_3\geq 1}\frac{(i h)^{n_1}(i A)^{n_2}(i A^*)^{n_3}}{n_1!n_2!n_3!}
 \left\langle f^{n_1}\Psi^{n_2}\Psi^{*n_3}\right\rangle_c\nonumber\\
 =-\frac{h^2G}{2}-AA^*\Pi-i hAA^*K+\ldots,
\end{eqnarray}
where $\Pi=\langle\Psi\Psi^*\rangle_c$ is the propagator. $K=\langle f\Psi\Psi^*\rangle_c$ is the vertex representing the
particle-field interaction,
 $$
\epsfig{file=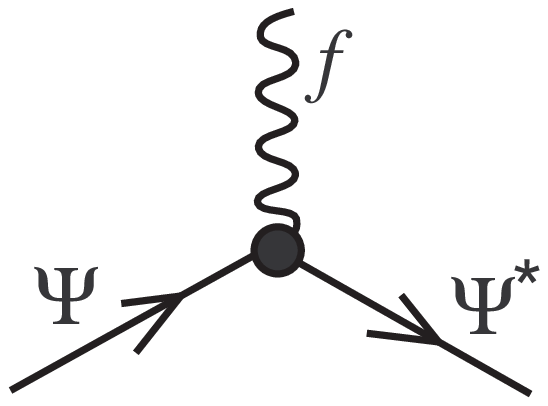, height=1.4cm},
 $$
where the wiggled edge corresponds to electric field $f$, and edges with arrows to complex fields $\Psi$ and $\Psi^*$.
Higher order cumulants does not survive after variation as the interaction part in the partition function (\ref{f12}) does
not include them. The presence of $K$ vertex can be included by means of diagrams. All diagrams that contain either first
order vertices or loops cancels. The simplest remaining diagram contains two $K$ vertices:
\begin{equation}\label{fdiagr}
 -\frac{1}{2}
 \raisebox{3pt}
  {
 \begin{picture}(30,0)
  \thicklines
  \put(15,0){\circle{20}}
  \put(5,0){\circle*{4}}
  \put(25,0){\circle*{4}}
  \bezier{0}(5,0)(7,3)(9,0)
  \bezier{0}(9,0)(11,-3)(13,0)
  \bezier{0}(13,0)(15,3)(17,0)
  \bezier{0}(17,0)(19,-3)(21,0)
  \bezier{0}(21,0)(23,3)(25,0)
 \end{picture}
  }
 =\frac{K^2}{2G\Pi\Pi}.
\end{equation}
Ignoring higher order diagrams, the average becomes
\begin{eqnarray}
 \langle\ln\langle\delta(f-\varphi)\delta(\Psi-\Theta)\delta(\Psi^*-\Theta^*)\rangle_{f,\Psi,\Psi^*}\rangle_{\varphi,\Theta,\Theta^*}\nonumber\\
 = -\frac{1}{2}(\ln(2\pi G)+1)-(\ln(-i 2\pi\Pi)+1) +\frac{K^2}{2G\Pi\Pi}. \label{dep8}
\end{eqnarray}
The next diagrams containing four vertices $K$ are not included. The accurate integration of the single integral (\ref{f1}),
containing only mentioned cumulants $G$, $\Pi$, and $K$ in the interaction part, shows that the sum of all diagrams gives
$\frac{1}{2}\ln\left(1+\frac{K^2}{G\Pi\Pi}\right)$, which agrees with (\ref{dep8}) at small $K$. However, the
interpretation of the result is non-trivial for many-dimensional integral without the use of diagram technique. $K$ is
approximately proportional to the charge of a particle $q_j$. Therefore, the higher order diagrams as (\ref{fdiagr}) can be
neglected considering the Coulomb interaction as a small perturbation. Therefore, the partition function of fermion plasma
is
\begin{eqnarray}\label{statsum0}
  \ln\mathcal{Z}&=&const+\frac{1}{2}\sum_{\mathbf{k},\omega}\left(\ln\frac{G_{\mathbf{k},\omega}}{V_k}+1-\frac{G_{\mathbf{k},\omega}}{V_k} \right)\nonumber\\
  &-&\sum_{j,\mathbf{k},\omega}\left(\ln \Pi_{j,\mathbf{k},\omega}+1-\alpha_{j,\mathbf{k},\omega}\Pi_{j,\mathbf{k},\omega}\right)\nonumber
  \\&+&\frac{i \sqrt{\beta}}{\sqrt{V}} \sum_{j,\mathbf{k},\omega,\mathbf{q},\chi}q_jK_{j,\mathbf{k},\omega,\mathbf{q},\chi}\\
  &+&\sum_{j,\mathbf{k},\omega,\mathbf{q},\chi}\frac{K_{j,\mathbf{k},\omega,\mathbf{q},\chi}^2}{2G_{\mathbf{k},\omega} \Pi_{j,\mathbf{q},\chi}
 \Pi_{j,\mathbf{k}+\mathbf{q},\omega+\chi}}\nonumber,
\end{eqnarray}
where $G_{\mathbf{k},\omega}=\langle|\varphi_{\mathbf{k},\omega}|^2\rangle$ is  cumulant of electric field,
$\Pi_{j,\mathbf{k},\omega}=\langle|\Psi_{j,\mathbf{k},\omega}|^2\rangle$ - propagator, and
$K_{j,\mathbf{k},\omega,\mathbf{q},\chi} =\left\langle\varphi_{\mathbf{k},\omega} \Psi_{j,\mathbf{q},\chi},
 \Psi_{j,\mathbf{k}+\mathbf{q},\omega+\chi}^*\right\rangle_{\varphi,\Psi,\Psi^*}$ - cumulant of field-par\-tic\-le
interaction. The constant can be readily determined by comparison with an ideal system, where
$\ln\mathcal{Z}=\sum_{j,\mathbf{k}}\ln[1+e^{-c_{j,\mathbf{k}}^{id}}]$. The first sum in the partition function corresponds
to contribution of field, the second one to noninteracting fermions, while the last ones are due to fermion-field
interaction. The variation of $K_{j,\mathbf{k},\omega,\mathbf{q},\chi}$ gives
\begin{equation}
 K_{j,\mathbf{k},\omega,\mathbf{q},\chi}=-\frac{i q_j\sqrt{\beta}}{\sqrt{V}}G_{\mathbf{k},\omega} \Pi_{j,\mathbf{q},\chi}
 \Pi_{j,\mathbf{k}+\mathbf{q},\omega+\chi}
\end{equation}
Thus, the diagram in (\ref{fdiagr}) is the only one which is proportional to the square of the charge assuring that
self-interaction part contracts exactly in sect. \ref{cmplxfld}. The functions $G_{\mathbf{k},\omega}$ and
$\Pi_{j,\mathbf{k},\omega}$ should be obtained from the minimum of the thermodynamic potential, too.

\section{Approximation of fermion propagator}
The correlation function $G_{\mathbf{k},\omega}$ is real but propagator $\Pi_{j,\mathbf{k},\omega}$ - complex-valued. It
seems that the main difficulties are connected with $\Pi_{j,\mathbf{k},\omega}$. Therefore, we will approximate the
propagator so that it depends on $\omega$ in the same way as reciprocal of $\alpha_{j,\mathbf{k},\omega}$ does:
\begin{equation}\label{ap3}
 \Pi_{j,\mathbf{k},\omega}=\frac{1}{N(e^{i\omega\varepsilon}-1)-i \pi+c_{j,\mathbf{k}}},
\end{equation}
where the real function $c_{j,\mathbf{k}}$ does not depend on $\omega$. If we try to give some physical meaning for this
function, it can be considered as self-energy. The variational principle is valid also for the new function
$c_{j,\mathbf{k}}$, which may be further simplified. The density of $j$-type particles is
\begin{equation}
 n_j=\frac{1}{V\beta}\frac{\partial\ln\mathcal{Z}}{\partial
 \mu_j}=-\frac{1}{V}\sum_{\mathbf{k},\omega}\Pi_{j,\mathbf{k},\omega}=\frac{1}{V}\sum_\mathbf{k}n_{j,\mathbf{k}},\label{conc}\\
\end{equation}
\begin{equation}
 n_{j,\mathbf{k}} =-\sum_\omega\Pi_{j,\mathbf{k},\omega} =-\sum_\omega\frac{1}{N(e^{i\omega\varepsilon}-1)-i \pi+c_{j,\mathbf{k}}}.\label{sum}
\end{equation}
\begin{figure}
 \mbox{\epsfig{file=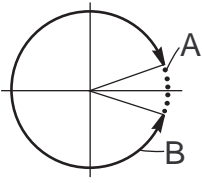,width=3cm}}
 \caption{Summation over $\omega$}
 \label{sumfig}
\end{figure}
The summation can be represented by a set of $N$ (see (\ref{t7b})) points distributed along a circle as shown in
Fig.~\ref{sumfig}. Let us divide the circle in two arcs: $A$, where $|\omega\varepsilon|\ll 1$; and $B$. Discrete summation
is necessary in the arc $A$. The arc includes infinite number of points at $N\rightarrow\infty$. If the angle of this arc
is sufficiently small, the propagator is approximately
$\Pi_{j,\mathbf{k},\omega}\approx(i\pi(2n-1)+c_{j,\mathbf{k}})^{-1}$. In the arc $B$, the summation can be safely replaced
by integration over $\phi=\omega\epsilon$. The integration limits are $-\pi$ and $\pi$, when the angle of the arc $A$ is
small. The propagator in the arc $B$ is $\Pi_{j,\mathbf{k},\omega}\approx(N(e^{i \phi}-1))^{-1}$. Hence, one obtains the
well known form of Fermi-Dirac distribution:
\begin{eqnarray}\label{sum1}
 n_{j,\mathbf{k}}&=&-\frac{1}{2\pi}\int_{-\pi}^\pi\frac{d\phi}{e^{i
 \phi}-1}-\sum_{n=-\infty}^{\infty}\frac{1}{i\pi(2n-1)+c_{j,\mathbf{k}}}\nonumber\\
 &=&\frac{1}{1+e^{c_{j,\mathbf{k}}}},
\end{eqnarray}
It is a consequence of approximation (\ref{ap3}). The integration of (\ref{sum1}) yields the sum of logarithm:
\begin{equation}\label{complex1}
 \sum_\omega\ln\Pi_{j,\mathbf{k},\omega}=-\ln[1+e^{-c_{j,\mathbf{k}}}]+const.
\end{equation}
The last sum of the partition function (\ref{statsum0}) contains the product of the propagators. The single sum of the
product at small $|\omega|$ is
\begin{eqnarray}\label{log}
 \sum_\chi\Pi_{j,\mathbf{q},\chi}\Pi_{j,\mathbf{q}+\mathbf{k},\chi+\omega} =\frac{n_{j,\mathbf{q}+\mathbf{k}}-n_{j,\mathbf{q}}}{i2\pi n+c_{j,\mathbf{q}+\mathbf{k}}-c_{j,\mathbf{q}}},\\
  |n|\ll N \nonumber,
\end{eqnarray}
because the integral along the arc $B$ (see Fig.~\ref{sumfig}) cancels. The double sum is a consequence of (\ref{sum}):
\begin{equation}\label{mnpp}
 \sum_{\omega,\chi}\Pi_{j,\mathbf{q},\omega}\Pi_{j,\mathbf{q}+\mathbf{k},\chi+\omega}=n_{j,\mathbf{q}}n_{j,\mathbf{q}+\mathbf{k}}.
\end{equation}
Thereafter, the partition function essentially simplifies
\begin{widetext}
\begin{eqnarray}\label{aftself}
 \ln\mathcal{Z}&=&\frac{1}{2}\sum_{\mathbf{k},n}\left(\ln\frac{G_{\mathbf{k},n}}{V_k}+1-\frac{G_{\mathbf{k},n}}{V_k} \right)+
 \sum_{j,\mathbf{k}}\left(\ln[1+e^{-c_{j,\mathbf{k}}}]+(c_{j,\mathbf{k}}-c_{j,\mathbf{k}}^{id})n_{j,\mathbf{k}}\right)\nonumber\\
 &+&\frac{\beta}{2V}\sum_{j,\mathbf{q},\mathbf{k}}q_j^2V_k\left(n_{j,\mathbf{q}}n_{j,\mathbf{q}+\mathbf{k}}-\sum_ng_{\mathbf{k},n}
 \frac{n_{j,\mathbf{q}+\mathbf{k}}-n_{j,\mathbf{q}}}
 {i 2\pi n+c_{j,\mathbf{k}+\mathbf{q}}-c_{j,\mathbf{q}}}\right),
\end{eqnarray}
\end{widetext}
where the summation index $n$ now runs from $-\infty$ to $\infty$. The last term does not contain singularities at $n=0$,
because $n_{j,\mathbf{k}}$, (\ref{sum1}), monotonously depends on $c_{j,\mathbf{k}}$. The partition function does not
contain divergent terms and so it is not necessary to introduce screened potentials. In the second sum, one recognises the
logarithm of thermodynamic probability
\begin{eqnarray}
 &&\ln[1+e^{-c_{j,\mathbf{k}}}]+c_{j,\mathbf{k}}n_{j,\mathbf{k}}\nonumber\\
&=&-n_{j,\mathbf{k}}\ln n_{j,\mathbf{k}}-(1-n_{j,\mathbf{k}})\ln(1-n_{j,\mathbf{k}}).
\end{eqnarray}
The variation of the correlation function of electric field yields
\begin{eqnarray}
 \Xi_{\mathbf{k},n}&\equiv&\frac{G_{\mathbf{k},n}-V_k}{G_{\mathbf{k},n}}\\
 &=&-\frac{\beta V_k}{V}\sum_{j,\mathbf{q}}q_j^2\frac{(n_{j,\mathbf{q}+\mathbf{k}}-n_{j,\mathbf{q}})(c_{j,\mathbf{k}+\mathbf{q}}-c_{j,\mathbf{q}})}
 {(2\pi n)^2+(c_{j,\mathbf{k}+\mathbf{q}}-c_{j,\mathbf{q}})^2}\nonumber.\label{extrem}
\end{eqnarray}
This relation proves the limit $\lim_{|\omega|\rightarrow\infty}G_{\mathbf{k},\omega}=V_k$ necessary for exclusion of
self-interaction in (\ref{t7a}). Subsequently, the partition function depends only on one function $c_{j,\mathbf{k}} $:
\begin{eqnarray}\label{ap20}
 \ln\mathcal{Z}&=&\frac{1}{2}\sum_{\mathbf{k},n}\left(\Xi_{\mathbf{k},n}-\ln[1+\Xi_{\mathbf{k},n}]\right)\nonumber\\
 &+&\sum_{j,\mathbf{k}}\left(\ln[1+e^{-c_{j,\mathbf{k}}}]+(c_{j,\mathbf{k}}-c_{j,\mathbf{k}}^{id})n_{j,\mathbf{k}}\right)\\
 &+&\frac{\beta}{2V}\sum_{j,\mathbf{q},\mathbf{k}}q_j^2V_k n_{j,\mathbf{q}}n_{j,\mathbf{q}+\mathbf{k}}\nonumber.
\end{eqnarray}
The variation of $c_{j,\mathbf{k}}$ seems more complicated. Hence, the restriction of $c_{j,\mathbf{k}}$ in a certain class
of real functions is necessary. This function may not be smooth at zero temperature, similarly, as the spectrum of some
substance becomes sharper decreasing the temperature. Note, that one can in principle obtain the ground state energies
for a given molecule performing the limit $T\rightarrow 0$ in the thermodynamic potential.\\
At low density limit, one obtains a classical Debye-H\"{u}ckel screening in position space: $G_0(\mathbf{x})=e^{-\alpha
x}/x$. The last term of (\ref{bozaft}) contains the product of distribution functions $n_{j,\mathbf{k}}$ and it is
negligible at low densities. Hence, the minimum of thermodynamic potential at low density limit gives the difference
\begin{equation}\label{cdev}
  c_{j,\mathbf{k}}-c_{j,\mathbf{k}}^{id}\approx-\frac{\beta q_j^2}{2}\kappa_D,
\end{equation}
where $\kappa_D$ is Debye-H\"{u}ckel screening radius. This difference does not depend on the wave vector $\mathbf{k}$. It
suggests that the presence of Coulomb interaction in the system is accountable by scaling of the chemical potential.

\subsection{Boson propagator} The function $\alpha_{j,\mathbf{k},n}$, (\ref{t11}), for bosons differs by $-i\pi$:
\begin{equation}
 \alpha_{j,\mathbf{k},n}=N(e^{i\frac{2\pi n}{N}}-1)+c_{j,\mathbf{k}}^{id}.
\end{equation}
The approximation of type (\ref{ap3}) leads to the well known form of Bose-Einstein distribution:
 $$
 n_{j,\mathbf{k}}=\frac{1}{2\pi}\int_{-\pi}^\pi\frac{dz}{e^{i
 z}-1}+\sum_{n=-\infty}^{\infty}\frac{1}{i 2\pi n+c_{j,\mathbf{k}}}=\frac{1}{e^{c_{j,\mathbf{k}}}-1}.
 $$
The partition for bosons can be obtained similarly as for fermions keeping in mind the comments for boson particles in
section \ref{cmplxfld}. Finally, one gets the partition function
\begin{eqnarray}\label{bozaft}
 \ln\mathcal{Z}&=&\frac{1}{2}\sum_{\mathbf{k},n}\left(\Xi_{\mathbf{k},n}-\ln[1+\Xi_{\mathbf{k},n}]\right)\nonumber\\
 &+&\sum_{j,\mathbf{k}}\left(-\zeta_j\ln[1-\zeta_je^{-c_{j,\mathbf{k}}}]+(c_{j,\mathbf{k}}-c_{j,\mathbf{k}}^{id})n_{j,\mathbf{k}}\right)\nonumber\\
 &-&\frac{\beta}{2V}\sum_{j,\mathbf{q},\mathbf{k}}\zeta_jq_j^2V_k n_{j,\mathbf{q}}n_{j,\mathbf{q}+\mathbf{k}},
\end{eqnarray}
\begin{equation}\label{boen}
 n_{j,\mathbf{k}}=\frac{1}{e^{c_{j,\mathbf{k}}}-\zeta_j},
\end{equation}
where $\zeta_j=\pm 1$ is for bosons and fermions, respectively. This partition function can be applied for plasmas
consisting of both fermions and bosons, e.g., deuterium plasma.
\begin{figure}
    \mbox{\epsfig{file=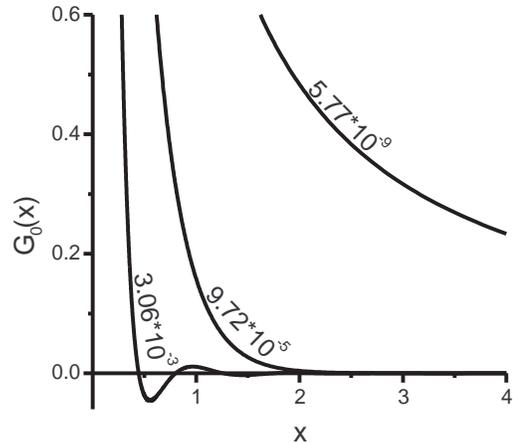, width=7cm}}
    \caption{Correlation function of electric field at $T=0.01$ and various densities for symmetrical plasma. Atomic units.}
    \label{cfe}
\end{figure}

\section{Symmetrical plasma}
\begin{figure}
    \mbox{\epsfig{file=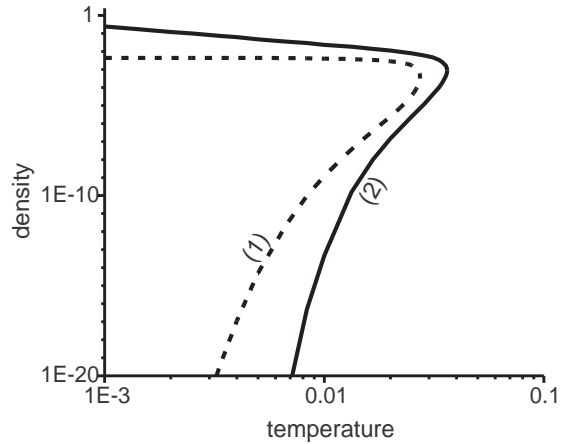, width=7.7cm}}
    \caption{Phase diagram of symmetrical plasma. Variational parameters: (1) - $\tilde{\mu}_j$; (2) - $\tilde{\mu}_j,\tilde{m}_j$. Atomic units.}
    \label{nT}
\end{figure}

Usually, only the simplest Coulomb systems are of theoretical interest, e.g., one-component, electron-hole (symmetrical
plasma), hydrogen, deuterium and helium plasmas. Here only the quantum symmetrical plasma is considered as an example, i.e.,
all species of particles are fermions. The symmetrical plasma is of interest because the controversial discussion exists
about the location of the critical point for its first order phase transition both in classical model of hard cores
\cite{Fisher} and quantum case \cite{Lehmann}. The unpolarised symmetrical plasma has four species $j$ of particles because
of two possible spin orientations. Despite the propagator has already been approximated in (\ref{ap3}), it is still
difficult to obtain the function $c_{j,\mathbf{k}}$ from the variation of thermodynamic potential (\ref{ap20}). Therefore,
only two parameters are variated for each specie of particle:
\begin{equation}\label{form}
 c_{j,\mathbf{k}}=\tilde{\eta}_j k^2-\beta\tilde{\mu}_j.
\end{equation}
Thus, a scaled chemical potential $\tilde{\mu}_j$ and an effective mass $\tilde{m}_j\equiv m_j\eta_j/\tilde{\eta}_j$ are
chosen as variational parameters. The correlation function of electric field is calculated on the basis of (\ref{extrem})
for every $n$. For convenience, the atomic units are used, and temperature is measured in the units of energy (hartrees).
The symmetrical plasma can be either electron-positron plasma or more familiar electron-hole plasma. However, the number of
particles in both systems does not conserve due to annihilation of antiparticles and recombination of electrons and holes
in semiconductor. The phase transition of electron-hole plasma is established both theoretically, e.g., in \cite{Kraeft},
\cite{Lehmann}, \cite{Shumway} and experimentally \cite{Rice}.
\begin{figure}
    \mbox{\epsfig{file=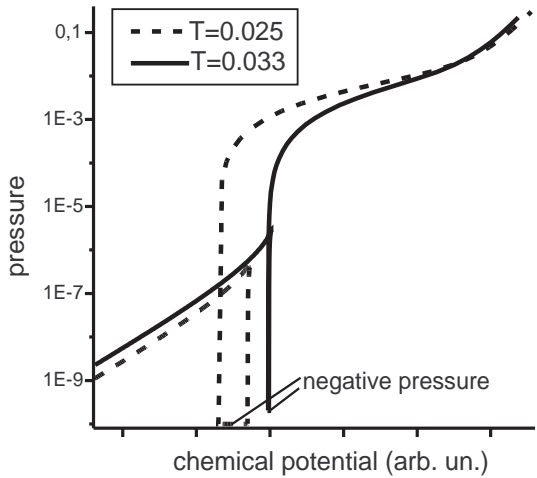, width=7.4cm}}
    \caption{Pressure isotherms vs. chemical potential for symmetrical plasma. The intersection point of curve corresponds to phase transition point.}
    \label{Pn}
\end{figure}
 \\It would be helpful to know what kind of the correlation function $G_{\mathbf{k},\omega}$,
(\ref{extrem}), one obtains for fermions, when $c_{j,\mathbf{k}}$ is chosen according to (\ref{form}). The correlation
function $G_{\mathbf{k},0}$ is of interest corresponding to an average with respect to imaginary-time since only terms with
$\omega=0$ are present in classical plasma. Fig.~\ref{cfe} shows the correlation function in position space $G_0(x)$ at
temperature $T=0.01$ and different densities. The correlation function transfers from screened behaviour, $G_0(x)\approx
V(x)=e^{-\alpha x}/x$, to oscillating one, when the Wigner-Seitz radius is comparable with the Bohr radius. The method does
not yield particle-particle distribution functions.\\
Since the partition function is linear with respect to the chemical potential $\mu_j$, it is better to fix the variational
parameter $\tilde{\mu}$ while $\mu_j$ follows from the system of variational equations. The effective mass monotonously
increases increasing the density. Thus, the presence of weak Coulomb interaction smoothes the step-like Fermi-Dirac
distribution function. The phase transition point can be found plotting the pressure vs. chemical potential (see
Fig.~\ref{Pn}) at temperatures below critical one. The intersection-point of isotherm corresponds to the coexistence of two
phases in accordance with phase equilibrium condition. Loop with negative pressure corresponds to unstable densities. The
obtained phase diagram of symmetrical plasma is shown in Fig.~\ref{nT}. The critical temperature is considerably lower, when
purely quantum terms $G_{\mathbf{k},n}$ with $n\neq 0$ are neglected. Obtained critical parameters of non-annihilating
electron-positron plasma are $T_c=11500 K$, $P_c=1.6\cdot10^8 Pa$, $n_c=5.7\cdot10^{27} m^{-3}$ (for curve (2)). The
critical temperature appears to be slightly higher than the corresponding quantum statistical result in \cite{Lehmann}:
$T_c=7186 K$, while the density is approximately the same. The hydrogen-like bound states are introduced directly in
\cite{Lehmann}, while the ground state energy in the presented functional integral model appears implicitly with slightly
different value at $T\rightarrow 0$, where the method does not work.
One notices in Fig.~\ref{nT} that the equilibrium concentration of the dense phase (curve (2)) is inversely proportional to
temperature. That would lead to collapse of symmetrical fermion plasma at zero temperature. However, the partition function
(\ref{ap20}) is, of course, not valid for strongly coupled plasmas. Simulations made by restricted path integral method
\cite{Shumway} show that such phenomena as Bose condensation of excitons and biexcitons takes place in dense electron-hole
plasma.

\section{Conclusions}
The paper is devoted mainly for elaboration of variational methodology that helps to investigate the quantum Coulomb systems
with Feynman path integral technique. The principal points the proposed scheme are:
\begin{enumerate}
  \item The representation of the partition function of fermion gas by functional integrals is obtained using Feynman path
integral and formally introduced integration over complex field. This integral contains only third order vertices that
simplifies further use of diagram technique.
  \item The form of the functional integral differs essentially for Fermi and Bose systems. For example, the integration
  over complex field for fermions stands in denominator, while for bosons - in numerator. Such a mathematically formal representation solves the problem
  of symmetrisation.
  \item The self-interaction part is excluded using modified Coulomb interaction potential at small distances and
  scaled chemical potential. The limit to accurate Coulomb interaction potential is performed in final expressions.
  \item The algorithm for expression of thermodynamic potential by average values cumulants of real and complex fields is
  elaborated. Those values of cumulants can be further found from the minimum of thermodynamic potential.
  \item The phase diagram and critical point are obtained using the simplest approximation of
  thermodynamic potential. For this reason, variation only of chemical potential
  and effective mass for correlation functions of particles is performed, while
  the minimisation in respect to cumulant of electric field is maintained exactly.
\end{enumerate}

\bibliography{varxxx}

\end{document}